\documentclass[%
 reprint,
 amsmath,amssymb,
 aps,
 prl,
]{revtex4-1}

\usepackage{graphicx}
\usepackage{dcolumn}
\usepackage{bm}

\begin{document}

\preprint{APS/123-QED}

\title{Granular-like behavior of molecular flow in constricted nanopores}

\author{Y. Magnin}
\email{magnin@mit.edu}
\affiliation{MIT, Energy Initiative, Massachusetts Institute of Technology, Cambridge, MA, United States}
\author{N. Chanut}
\affiliation{MIT, Energy Initiative, Massachusetts Institute of Technology, Cambridge, MA, United States}
\author{D.A. Pantano}
\affiliation{Total Marketing \& Services, Chemin du Canal, 69360 Solaize, France}
\author{E. Chaput}
\affiliation{Total E\&P Research \& D\'evelopement, CSTJF - Avenue Larribau, 64018 Pau Cedex, France}
\author{F-J. Ulm}
\affiliation{MIT, Energy Initiative, Massachusetts Institute of Technology, Cambridge, MA, United States}
\author{R. J.-M. Pellenq}
\affiliation{Multi-Scale Materials Science for Energy and Environment the MIT / CNRS / Aix-Marseille University Joint Laboratory at Massachusetts Institute of Technology, Cambridge, Massachusetts 02139, USA}
\altaffiliation[Also at ]{Department of Physics, Georgetown University, Washington DC, USA}
\author{E. Villermaux}
\affiliation{Aix Marseille Universit\'e, CNRS, Centrale Marseille, IRPHE UMR 7342, Marseille 13384, France}
\date{\today}

\begin{abstract}
The fluid flow through porous media is described by Darcy's law, while the fluid/wall interactions can be neglected. In nanopores, where adsorption dominates, Darcy's extension has been made, but approaches able to describe flows in mesopores are still lacking. We show here that molecular flows through nano-constrictions is well described by Berverloo's law, predicting the mass flow rate of granular material through macro-apertures. This molecular and granular analogy, allows to derive a relationship providing a theoretical framework for the molecular flow in disordered mesoporous systems.
\end{abstract}

\maketitle
The flow of matter through an aperture is a common situation, spanning a broad range of length scales and applications. At the human scale, it is typically found in sand flowing in an hourglass, seeds discharge from a silo\cite{pouliquen,weinhart2016}, traffic jam on merging lanes \cite{nagel,helbing}, or even in humans and animals collective motion through a narrow exit \cite{pastor,garcimartin}. At the micrometer scale, it can be found in the human body such as the blood flow in arteries \cite{gutterman}, in fluid filtration systems \cite{chang,ang,linkhorst}, or in microfluidic devices \cite{higgins,genovese,wyss,macdonald}. In these situations, flows can be described by the long-established laws of fluid or granular materials, depending on the nature of the flowing particles. However, the extension of these macroscopic laws in nano-channels is not trivial due to the emergence of physical and chemical properties in such a confined space \cite{falk1}. The presence of constrictions at the nanometer scale is indeed typical in a wide range of applications involving disordered porous system, including but not limited to i) the hydrocarbon recovery from shale rocks \cite{berthonneau}, ii) electrical energy storage (Li-ion batteries \cite{lin}, supercapacitors \cite{borchardt}), iii) catalytic and adsorption-based separation processes \cite{mitchell,chanut} or iv) the molecular diffusion across membranes in biological systems \cite{zxu}.\\
\noindent The transport properties of a fluid in a macro-porous solid (defined as pore diameter $h\ >$ 500 \AA\ in the IUPAC classification \cite{rouquerol}) is well described by Darcy's law:
\begin{eqnarray}
v=-\frac{b}{\eta}\ \nabla P.
\label{darcy}
\end{eqnarray}
\begin{figure*}
\includegraphics[scale=0.30]{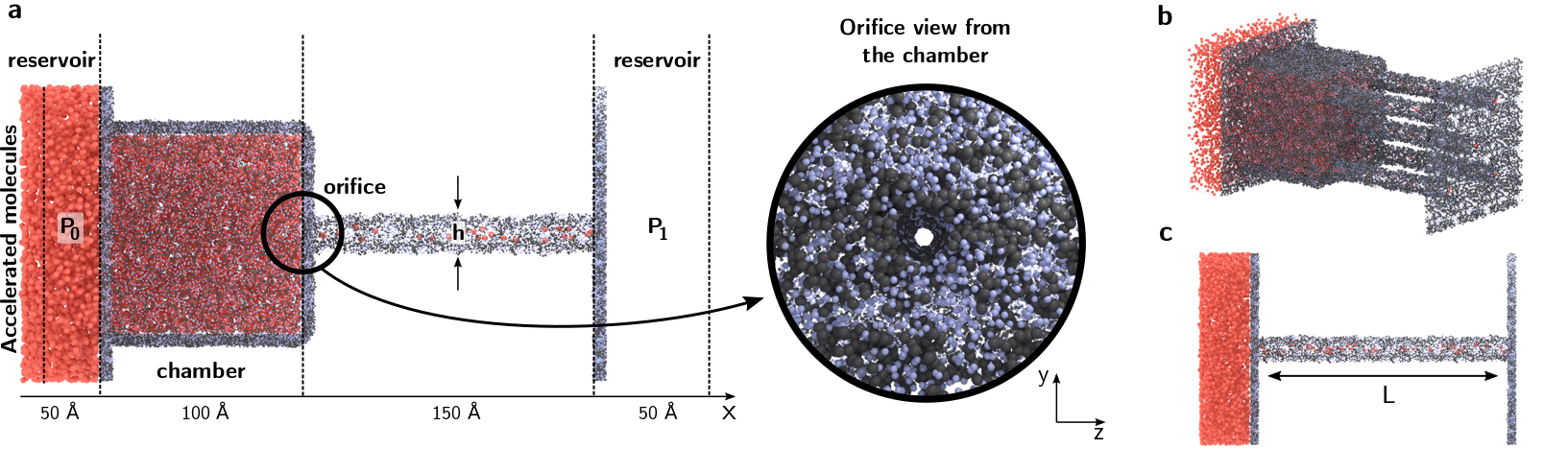}
\caption{\label{fig1} Atomistic structures designed to simulate the molecular flow, made of carbon (dark) and hydrogen atoms (blue). For constrictions, a chamber is connected to two reservoirs in $x$ direction. The left reservoir is filled by methane molecules (red) at a pressure $P_0$, while an empty reservoir is maintained at $P_1\sim 0$ on the right side of the structure. a. Synthetic constriction consisting in a chamber, connected to a pore of length $L$ and diameter $h$, forming an orifice at the connection between the two. b. Synthetic constriction connected to five pores of same diameters $h$=10\AA. c. Structure without constriction.}
\end{figure*}
With $v$ the fluid velocity through the porous material, $b$ the permeability of the porous medium, $\eta$ the fluid viscosity, and $\nabla P$ the pressure gradient. As the permeability is an intrinsic property of the porous medium, an inherent point to the application of Darcy's law is the constant viscosity of the fluid at given thermodynamic conditions. However, when a fluid is brought into contact with a solid, adsorption may occurs. It is defined as the increase in the fluid density close to the solid surface, due to the fluid-wall attractive interactions \cite{falk1}. When pores are sufficiently large, the adsorption is negligible and $\eta$ is equal to the one of the bulk. In that case, the transport is purely convective and equation \eqref{darcy} is applicable. When the pore size decreases, the adsorption can no longer be neglected, resulting in fluctuations of the density along the radial direction of the pore. In that situation, here referred to as the mesoscale, $\eta$ depends on the pore size and molecules experience a transport that is both diffusive and convective \cite{frentrup}. For pores below a few molecular diameters \cite{sholl}, here referred to as nanopores, the high confinement leads to an uni-dimensional transport along the pore, where molecules behave individually and where their collective interactions can be neglected \cite{valuilin,falk2_ar}. In such conditions, the concept of fluid viscosity - that is in essence multi-body - becomes meaningless. The change in the transport properties resulting from a strong confinement at the nano- and mesoscale makes Darcy's law inapplicable \cite{travis,falk2_ar,bhatia,sholl}.\\
To address the question of the transport of hydrocarbon in shale rocks at the nanoscale, Falk et al.\cite{falk2_ar} have recently shown that equation \eqref{darcy} can be rewritten such as:
\begin{eqnarray}
v=-B\ \nabla P,\ \ \ \ \ \ B=\frac{D_c}{\rho\ k_BT},\ \ \ \ \ \ D_c\ \sim\ D_s.
\label{ndarcy}
\end{eqnarray}
In equation \eqref{ndarcy}, $B$ is the permeance, defined as the ratio of the collective diffusion coefficient $D_c$, with the fluid density $\rho$ and the thermal energy $k_BT$. In such a strong confinement situation, the self and collective diffusion coefficients - respectively $D_s$ and $D_c$ - are almost equal due to the lack of collective interactions \cite{falk2_ar}. This equation that we will call nano-Darcy for the sake of clarity, has been successfully applied to describe the transport of hydrocarbons in fully nanoporous systems (pore diameters $<20$\AA\ \cite{rouquerol}). However, it has never been questioned at the mesoscale, where molecular collective interactions are partially recovered (pore diameters between 20 and 500\AA\ \cite{rouquerol}).\\ 
The molecular transport is generally considered as a fluid mechanics problem, hence no links with the flow of granular materials have been made so far. In contrast to fluids, the discharge rate of grains through an aperture does not depend on the container's size neither on the height of the grains layer above the constriction \cite{aguirre,pouliquen}. For granular materials, the athermal mass flow rate therefore depends solely on the density of the granular bed, and on the geometrical considerations of the system as expressed by Berverloo's law \cite{beverloo}:
\begin{equation}
\label{eq:1}
W\ =\ \rho v S\ =\ C\ \rho\sqrt{g}\ (h-k\sigma)^{5/2}.
\end{equation}
In equation \eqref{eq:1}, $W$ is the mass flow rate, $\rho$ is the grain density, $S$ the pore surface, $h$ the aperture diameter, $\sigma$ the particle diameter, and $g$ the earth gravity. $C$ and $k$ are empirical parameters linked to the grain friction on the container walls and the grain aspect ratio, respectively \cite{nedderman}. In this expression, $W$ reads as the product of an exit section $h^2$ (in three dimensions) times a free fall velocity $\sqrt{gh}$ on a length corresponding to the aperture dimension $h$, when the packed grains are pushed by a body force $mg$ \cite{hagen} (with $m$ the mass of the grain). Historically proposed for non-brownian flowing grains at the macroscopic scale and widely admitted in this context, the relevance of the Beverloo's law has however never been questioned at the nano- and mesoscale for a molecular confined fluid whose behavior is driven by adsorption thermodynamics. For that purpose, synthetic carbonaceous atomistic structures have been designed and the flow of methane molecules through these has been studied by means of non-equilibrium molecular dynamics (NEMD), Fig.\ref{fig1}.
\begin{figure*}
\includegraphics[scale=0.29]{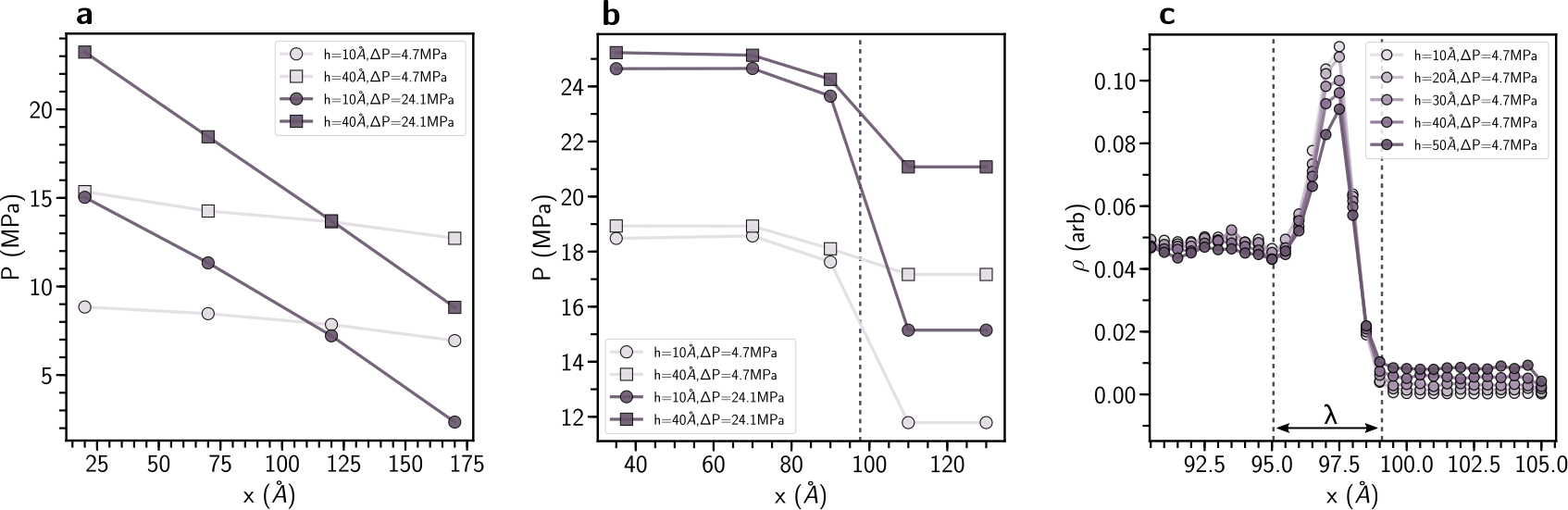}
\caption{\label{fig2} Evolution of the fluid pressure and density along the flowing direction $x$. a. Linear pressure decrease along a pore of length $L$ without constriction. b. Non-linear pressure decrease in a system with a constriction. In (a) and (b), circles correspond to a pore size $h=$10\AA, while squares correspond to $h=$40\AA. The light gray plots correspond to $\Delta P=$4.7MPa and the dark gray plots to $\Delta P$=24.1MPa. The dashed line in (b) indicates the position of the aperture. c. Fluid density at the vicinity of the aperture. $\lambda$ is a typical length scale of high molecular concentration.}
\end{figure*}
\noindent Numerical simulations are performed in synthetic systems consisting in a chamber of 100$\times$100$\times$100 \AA$^3$\, connected by a pore of length $L$=150\AA\ and diameter $h$ (ranging from 10 to 50 \AA), forming an orifice at the connection between the two. Each side of the structure is connected to a reservoir, the pressure difference ($\Delta P=P_0-P_1$) sustaining a flow of methane molecules through the system, Fig.\ref{fig1}a. Simulations are performed for different $\Delta P$ (ranging from $\sim$1 to 25 MPa), temperatures (300 and 400$K$), pore diameters, and connectivities $\xi$ (one or five branching pores, Fig.\ref{fig1}b). The range of pressure and temperature has been chosen to cover both gas and supercritical states of confined methane molecules (for pore diameters above 2 nm as the distinction is not clear at the nanoscale) \cite{wu}. The simulations have also been performed in systems without constrictions (pore length $L$=200\AA), for comparison, Fig.\ref{fig1}c. It is important to note that a constriction is defined as the chamber and its connected pore(s), excluding the two reservoirs. Details about simulations are presented in Supplemental Material.
\noindent In a pore of length $L$ without constriction, the pressure difference $\Delta P$ in the system originates a continuous pressure decay along the $x$ direction, Fig.\ref{fig2}a, corresponding to a pressure gradient $-\nabla P=\Delta P/L$. In the pore size range considered here, we observe that $\nabla P$ is independent of the pore diameter $h$. In contrast, an abrupt pressure drop is observed at the orifice in presence of constrictions, while molecules travel in the remaining pore volume with almost no pressure loss \cite{shen}, Fig.\ref{fig2}b. Interestingly, this behavior is reminiscent of the flow of granular materials through constrictions at the macroscale \cite{pouliquen}. The pressure drop at the orifice corresponds to a jamming phenomenon, confirmed by a peak of the molecular density $\rho$, spanning over a distance $\lambda$ and defining a pressure gradient such as $-\nabla P=\Delta P/\lambda$, Fig.\ref{fig2}c. It is noteworthy that the length $\lambda$ is almost independent of $\nabla P$ and $h$. In the present case of methane adsorbed on a carbonaceous material, we note that $\lambda\sim \sigma$, indicating that the increased density at the orifice is due to a monolayer of methane adsorbed on the surface. Further simulations performed with a repulsive inter-atomic potential does not show this peak of density at the orifice. However, although less marked, the pressure drop is also observed suggesting that this granular-like flow does not originate from adsorption effects alone (see Fig.3a,b in Supplemental Material). In addition, we also observe a linear increase of the molecular density in the system as a function of the pressure gradient. This highlights a slight compressibility of the fluid in the pressure range investigated (see Fig.4 in Supplemental Material).
\noindent The molecular velocity $v$ as a function of the pore diameter and the pressure gradient has then been determined from the NEMD at 400K in systems with and without constrictions. Without constriction, $v$ scales linearly with the pressure gradient (see Fig.5a in Supplemental Material), as expected from the linear response theory, similarly to (nano-)Darcy's laws. The same behavior is observed in presence of constrictions for moderate $-\nabla P=\Delta P/\lambda$, corresponding to $\Delta P <$ 5MPa. However, for larger pressure gradient, $v$ slightly deviates from this linear behavior due to moderate fluid compressibility effects (see Fig.5b in Supplemental Material). This weak deviation will be neglected in the rest of the paper and we approximate $v$ as a linear function of $\Delta P/\lambda$. An extension to the non-linear case is proposed in Supplemental Material.\\
Due to the observed similarity between molecular and granular flow, we then made use of Beverloo's law to model the transport of methane in constricted systems. It is however necessary to adapt the equation \eqref{eq:1}, as the force driving the molecular flow at the nanoscale differs from the body force pushing the grains in a macroscopic system. It is done by replacing $g$ with $\Delta P/(\lambda \rho)$. The mass flow rates in constricted systems have thus been determined as a function of $h$, and for different $\Delta P/\lambda$, with a good agreement when compared to the theoretical values obtained from equation \eqref{eq:1}, Fig.\ref{fig3}a. Significantly, this highlights a granular-like behavior of the molecular flow in presence of meso-constrictions, which, to the best of our knowledge, has never been shown before.
\begin{figure*}
\includegraphics[scale=0.24]{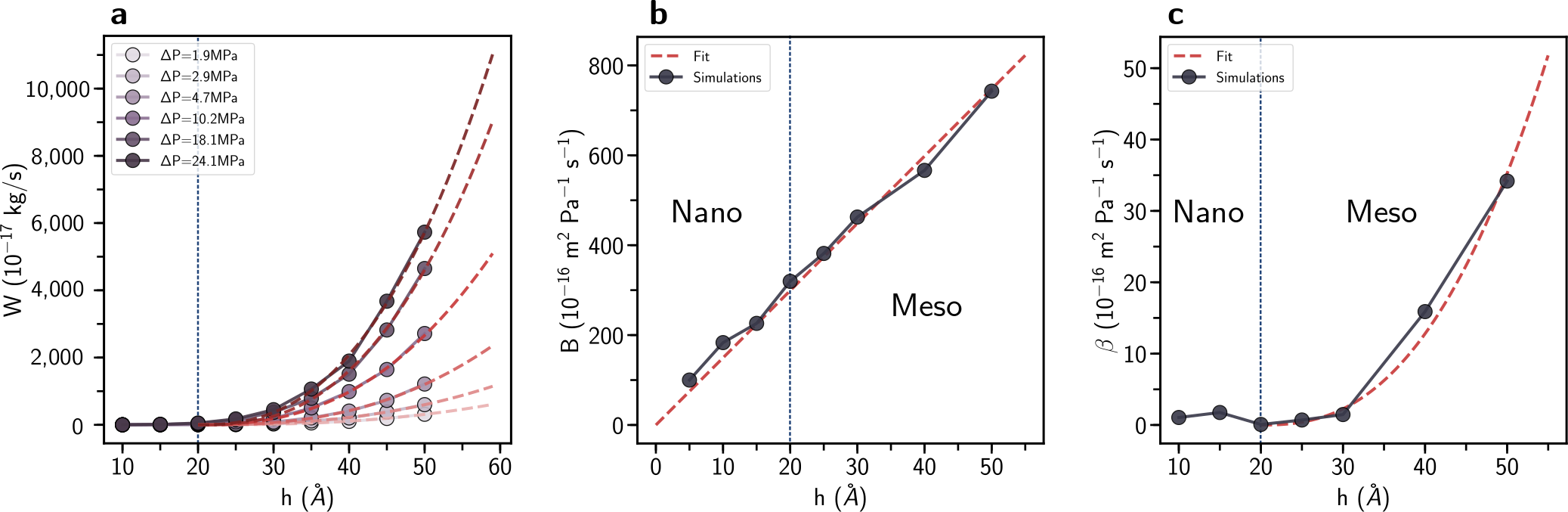}
\caption{a. Mass flow rate determined by numerical simulations as a function of the pore diameter $h$ for different $\Delta P$ (circles). The dashed lines correspond to the fit from the Beverloo law. b. Permeance $B$ as a function of the pore diameter in a system without constrictions. c. Permeance $\beta$ as a function of the pore diameter in a constricted system. The red dashed lines correspond to the fit of the simulations.}
\label{fig3}
\end{figure*}
\noindent The fitting procedure is performed by adjusting the two parameters $k$ and $C$ in equation \eqref{eq:1}. On the one hand, $k$ is found to be constant $k\sim5.4$, similar to the case of macroscopic granular media (see Fig.6a in Supplemental Material). In equation \eqref{eq:1}, $W\propto(h-k\sigma)^{5/2}$, implying that it is no longer define for $h< k\sigma$. Thus, for methane with a diameter $\sigma=3.7$\AA, the equation \eqref{eq:1} does not allow any prediction for $h\ \lesssim\ 20$\AA. In this situation, the formalism used here predicts a molecular clogging, while simulations show a finite flow in a pure diffusive regime, indicating the limit of validity of the approach. On the other hand, $C$ is found to scale as $\sqrt{\Delta P/\lambda}$ (see Fig.6b in Supplemental Material). This is a difference with the situation prevailing in macroscopic granular media, where $C$ is generally considered as a constant. Such distinction can be reasonably attributed to thermodynamic effects, which are not playing a role in granular materials discharge.\\
The evolution of the velocity as a function of $h$ has then been studied for the two considered topologies at 400K and pressures ranging from 5 to 24MPa (methane in its supercritical state). As already mentioned, $v\ \propto\ \nabla P$ corresponding to a Darcy-like flow in systems with and without constriction. However, while $v$ scales almost linearly with $h$ in straight pores, it is found to scale as a power law when molecules are flowing through a constriction, (see Fig.7a,b, respectively in Supplemental Material). As adsorption is taken into account in both cases, such a difference highlight the critical impact of the topology on the mechanisms governing the molecular flow. Such a behavior is also confirmed for methane in a gas state (at 300K and $\Delta P\sim$1MPa \cite{wu}), highlighting a similar transport mechanism in both gas and supercritical states (see Fig.8a,b in Supplemental Material).\\
In order to propose an analytical expression for $W$ in a constricted system, we reformulate equation \eqref{ndarcy} based on the velocity obtained from the simulations. To do so, the permeance $B$ in equation \eqref{ndarcy}, that is found to scale linearly as a function of $h$, Fig.\ref{fig3}.b, is replaced by a permeance for constricted systems $\beta\ \propto\ h^{5/2}$, Fig.\ref{fig3}c. This difference can be readily explained by the different nature of the molecular flow in systems with and without constrictions. We note that $\beta$ is found to be independent of $\Delta P/\lambda$ and can be fitted as $\beta\ =\ C_0\ (h-k\sigma)^{5/2}$, with C$_0$ a constant. Coupling equations \eqref{ndarcy} and \eqref{eq:1}, and replacing $B$ by $\beta$ then allows to formulate $C$ and $W$ such as,
\begin{eqnarray}
\label{eq:Cdef}
C\ &=&\ C_0\ S\ \sqrt{\rho_{eq}\ \frac{\Delta P}{\lambda}},\nonumber \\
W\ &=&\ C_0\ S\ \rho_{eq}\ (h-k\sigma)^{5/2}\ \frac{\Delta P}{\lambda}.
\end{eqnarray}
\noindent Thus, we show that the Beverloo's law can be used to formulate a linear relationship of $W\ \propto\ \Delta P/\lambda$ with a permeance that is shown to only depends on topological parameters of the porous structure. The numerical values of these parameters can be found in Supplemental Material. Moreover, the effect of the pore connectivity $\xi$ can be is easily incorporated in equation \eqref{eq:Cdef}. Interestingly, it is shown that if $\xi=n$ (with $n>1$), the overall flow in a system containing $n$ branched pores is exactly $n$ times the one of a system made of only one branched pores (see Fig.9 in Supplemental Material). This result highlights the versatility of the approach developed for its application in realistic systems.\\
As previously mentioned, equation \eqref{eq:Cdef} has been derived for incompressible fluids as it will be the case encountered in most applications.\\

The successful application of Beverloo's law in presence of constrictions at the nanoscale make an unexpected parallel between molecular and granular flow, that we coin as "molecular granularity". Some differences between the athermal flow of grains and the thermodynamically-driven molecular transport must however be taken into account. First, the compressible nature of the fluid at extreme thermodynamic conditions, in contrast to the incompressibility of grains. Second, the adsorption of the fluid molecules on the pore wall. The effect of molecular interactions between the fluid and the pore walls has been investigated for both attractive and repulsive interactions. It is observed, for a given topology, that the strength of interactions has an effect on the magnitude of the molecules density at the constriction, and therefore on their velocity. However, the pressure drop at the orifice, characteristic of granular materials flow, is still present. This shows that the thermodynamically-driven flow of molecules will be impacted by the strength of their interactions with the pore walls, but that the granular-like flow of molecules mainly originates from the porous topology. This analogy between granular and molecular flow further allowed us to establish a theoretical framework linking the mass flow rate with the pressure gradient, in a linear response theory. Due to the complexity of questions pertaining to molecular transport at the nanometer scale, this study propose a novel way of apprehending molecular transport in constricted systems through the use of granular physics, an axis unexplored until now.\\

\section*{Acknowledgments}
This work has been carried out within the support of CNRS and AMIDEX, the Aix-Marseille University foundation. We acknowledge funding from Total through the MIT/CNRS/AMU FASTER-Shale project. Y. Magnin also gratefully acknowledges TOTAL for the CPU time provided on their large super calculator PANGEA.\\


\begin{thebibliography}{34}%
\makeatletter
\providecommand \@ifxundefined [1]{%
 \@ifx{#1\undefined}
}%
\providecommand \@ifnum [1]{%
 \ifnum #1\expandafter \@firstoftwo
 \else \expandafter \@secondoftwo
 \fi
}%
\providecommand \@ifx [1]{%
 \ifx #1\expandafter \@firstoftwo
 \else \expandafter \@secondoftwo
 \fi
}%
\providecommand \natexlab [1]{#1}%
\providecommand \enquote  [1]{``#1''}%
\providecommand \bibnamefont  [1]{#1}%
\providecommand \bibfnamefont [1]{#1}%
\providecommand \citenamefont [1]{#1}%
\providecommand \href@noop [0]{\@secondoftwo}%
\providecommand \href [0]{\begingroup \@sanitize@url \@href}%
\providecommand \@href[1]{\@@startlink{#1}\@@href}%
\providecommand \@@href[1]{\endgroup#1\@@endlink}%
\providecommand \@sanitize@url [0]{\catcode `\\12\catcode `\$12\catcode
  `\&12\catcode `\#12\catcode `\^12\catcode `\_12\catcode `\%12\relax}%
\providecommand \@@startlink[1]{}%
\providecommand \@@endlink[0]{}%
\providecommand \url  [0]{\begingroup\@sanitize@url \@url }%
\providecommand \@url [1]{\endgroup\@href {#1}{\urlprefix }}%
\providecommand \urlprefix  [0]{URL }%
\providecommand \Eprint [0]{\href }%
\providecommand \doibase [0]{http://dx.doi.org/}%
\providecommand \selectlanguage [0]{\@gobble}%
\providecommand \bibinfo  [0]{\@secondoftwo}%
\providecommand \bibfield  [0]{\@secondoftwo}%
\providecommand \translation [1]{[#1]}%
\providecommand \BibitemOpen [0]{}%
\providecommand \bibitemStop [0]{}%
\providecommand \bibitemNoStop [0]{.\EOS\space}%
\providecommand \EOS [0]{\spacefactor3000\relax}%
\providecommand \BibitemShut  [1]{\csname bibitem#1\endcsname}%
\let\auto@bib@innerbib\@empty
\bibitem [{\citenamefont {Andreotti}\ \emph {et~al.}(2013)\citenamefont
  {Andreotti}, \citenamefont {Forterre},\ and\ \citenamefont
  {Pouliquen}}]{pouliquen}%
  \BibitemOpen
  \bibfield  {author} {\bibinfo {author} {\bibfnamefont {B.}~\bibnamefont
  {Andreotti}}, \bibinfo {author} {\bibfnamefont {Y.}~\bibnamefont {Forterre}},
  \ and\ \bibinfo {author} {\bibfnamefont {O.}~\bibnamefont {Pouliquen}},\
  }\href@noop {} {\emph {\bibinfo {title} {Granular media: between fluid and
  solid}}}\ (\bibinfo  {publisher} {Cambridge University Press},\ \bibinfo
  {year} {2013})\BibitemShut {NoStop}%
\bibitem [{\citenamefont {Weinhart}\ \emph {et~al.}(2016)\citenamefont
  {Weinhart}, \citenamefont {Labra}, \citenamefont {Luding},\ and\
  \citenamefont {Ooi}}]{weinhart2016}%
  \BibitemOpen
  \bibfield  {author} {\bibinfo {author} {\bibfnamefont {T.}~\bibnamefont
  {Weinhart}}, \bibinfo {author} {\bibfnamefont {C.}~\bibnamefont {Labra}},
  \bibinfo {author} {\bibfnamefont {S.}~\bibnamefont {Luding}}, \ and\ \bibinfo
  {author} {\bibfnamefont {J.~Y.}\ \bibnamefont {Ooi}},\ }\href@noop {}
  {\bibfield  {journal} {\bibinfo  {journal} {Powder technology}\ }\textbf
  {\bibinfo {volume} {293}},\ \bibinfo {pages} {138} (\bibinfo {year}
  {2016})}\BibitemShut {NoStop}%
\bibitem [{\citenamefont {Nagel}\ and\ \citenamefont {Paczuski}(1995)}]{nagel}%
  \BibitemOpen
  \bibfield  {author} {\bibinfo {author} {\bibfnamefont {K.}~\bibnamefont
  {Nagel}}\ and\ \bibinfo {author} {\bibfnamefont {M.}~\bibnamefont
  {Paczuski}},\ }\href@noop {} {\bibfield  {journal} {\bibinfo  {journal}
  {Physical Review E}\ }\textbf {\bibinfo {volume} {51}},\ \bibinfo {pages}
  {2909} (\bibinfo {year} {1995})}\BibitemShut {NoStop}%
\bibitem [{\citenamefont {Helbing}(2001)}]{helbing}%
  \BibitemOpen
  \bibfield  {author} {\bibinfo {author} {\bibfnamefont {D.}~\bibnamefont
  {Helbing}},\ }\href@noop {} {\bibfield  {journal} {\bibinfo  {journal}
  {Reviews of modern physics}\ }\textbf {\bibinfo {volume} {73}},\ \bibinfo
  {pages} {1067} (\bibinfo {year} {2001})}\BibitemShut {NoStop}%
\bibitem [{\citenamefont {Pastor}\ \emph {et~al.}(2015)\citenamefont {Pastor},
  \citenamefont {Garcimart{\'\i}n}, \citenamefont {Gago}, \citenamefont
  {Peralta}, \citenamefont {Mart{\'\i}n-G{\'o}mez}, \citenamefont {Ferrer},
  \citenamefont {Maza}, \citenamefont {Parisi}, \citenamefont {Pugnaloni},\
  and\ \citenamefont {Zuriguel}}]{pastor}%
  \BibitemOpen
  \bibfield  {author} {\bibinfo {author} {\bibfnamefont {J.~M.}\ \bibnamefont
  {Pastor}}, \bibinfo {author} {\bibfnamefont {A.}~\bibnamefont
  {Garcimart{\'\i}n}}, \bibinfo {author} {\bibfnamefont {P.~A.}\ \bibnamefont
  {Gago}}, \bibinfo {author} {\bibfnamefont {J.~P.}\ \bibnamefont {Peralta}},
  \bibinfo {author} {\bibfnamefont {C.}~\bibnamefont {Mart{\'\i}n-G{\'o}mez}},
  \bibinfo {author} {\bibfnamefont {L.~M.}\ \bibnamefont {Ferrer}}, \bibinfo
  {author} {\bibfnamefont {D.}~\bibnamefont {Maza}}, \bibinfo {author}
  {\bibfnamefont {D.~R.}\ \bibnamefont {Parisi}}, \bibinfo {author}
  {\bibfnamefont {L.~A.}\ \bibnamefont {Pugnaloni}}, \ and\ \bibinfo {author}
  {\bibfnamefont {I.}~\bibnamefont {Zuriguel}},\ }\href@noop {} {\bibfield
  {journal} {\bibinfo  {journal} {Physical Review E}\ }\textbf {\bibinfo
  {volume} {92}},\ \bibinfo {pages} {062817} (\bibinfo {year}
  {2015})}\BibitemShut {NoStop}%
\bibitem [{\citenamefont {Garcimart{\'\i}n}\ \emph {et~al.}(2015)\citenamefont
  {Garcimart{\'\i}n}, \citenamefont {Pastor}, \citenamefont {Ferrer},
  \citenamefont {Ramos}, \citenamefont {Mart{\'\i}n-G{\'o}mez},\ and\
  \citenamefont {Zuriguel}}]{garcimartin}%
  \BibitemOpen
  \bibfield  {author} {\bibinfo {author} {\bibfnamefont {A.}~\bibnamefont
  {Garcimart{\'\i}n}}, \bibinfo {author} {\bibfnamefont {J.~M.}~\bibnamefont
  {Pastor}}, \bibinfo {author} {\bibfnamefont {L.~M.}~\bibnamefont {Ferrer}},
  \bibinfo {author} {\bibfnamefont {J.~J.}~\bibnamefont {Ramos}}, \bibinfo
  {author} {\bibfnamefont {C.}~\bibnamefont {Mart{\'\i}n-G{\'o}mez}}, \ and\
  \bibinfo {author} {\bibfnamefont {I.}~\bibnamefont {Zuriguel}},\ }\href@noop
  {} {\bibfield  {journal} {\bibinfo  {journal} {Physical Review E}\ }\textbf
  {\bibinfo {volume} {91}},\ \bibinfo {pages} {022808} (\bibinfo {year}
  {2015})}\BibitemShut {NoStop}%
\bibitem [{\citenamefont {Gutterman}\ \emph {et~al.}(2016)\citenamefont
  {Gutterman}, \citenamefont {Chabowski}, \citenamefont {Kadlec}, \citenamefont
  {Durand}, \citenamefont {Freed}, \citenamefont {Ait-Aissa},\ and\
  \citenamefont {Beyer}}]{gutterman}%
  \BibitemOpen
  \bibfield  {author} {\bibinfo {author} {\bibfnamefont {D.~D.}\ \bibnamefont
  {Gutterman}}, \bibinfo {author} {\bibfnamefont {D.~S.}\ \bibnamefont
  {Chabowski}}, \bibinfo {author} {\bibfnamefont {A.~O.}\ \bibnamefont
  {Kadlec}}, \bibinfo {author} {\bibfnamefont {M.~J.}\ \bibnamefont {Durand}},
  \bibinfo {author} {\bibfnamefont {J.~K.}\ \bibnamefont {Freed}}, \bibinfo
  {author} {\bibfnamefont {K.}~\bibnamefont {Ait-Aissa}}, \ and\ \bibinfo
  {author} {\bibfnamefont {A.~M.}\ \bibnamefont {Beyer}},\ }\href@noop {}
  {\bibfield  {journal} {\bibinfo  {journal} {Circulation research}\ }\textbf
  {\bibinfo {volume} {118}},\ \bibinfo {pages} {157} (\bibinfo {year}
  {2016})}\BibitemShut {NoStop}%
\bibitem [{\citenamefont {Chang}\ \emph {et~al.}(2002)\citenamefont {Chang},
  \citenamefont {Le~Clech}, \citenamefont {Jefferson},\ and\ \citenamefont
  {Judd}}]{chang}%
  \BibitemOpen
  \bibfield  {author} {\bibinfo {author} {\bibfnamefont {I.-S.}\ \bibnamefont
  {Chang}}, \bibinfo {author} {\bibfnamefont {P.}~\bibnamefont {Le~Clech}},
  \bibinfo {author} {\bibfnamefont {B.}~\bibnamefont {Jefferson}}, \ and\
  \bibinfo {author} {\bibfnamefont {S.}~\bibnamefont {Judd}},\ }\href@noop {}
  {\bibfield  {journal} {\bibinfo  {journal} {Journal of environmental
  engineering}\ }\textbf {\bibinfo {volume} {128}},\ \bibinfo {pages} {1018}
  (\bibinfo {year} {2002})}\BibitemShut {NoStop}%
\bibitem [{\citenamefont {Ang}\ and\ \citenamefont {Elimelech}(2007)}]{ang}%
  \BibitemOpen
  \bibfield  {author} {\bibinfo {author} {\bibfnamefont {W.~S.}\ \bibnamefont
  {Ang}}\ and\ \bibinfo {author} {\bibfnamefont {M.}~\bibnamefont
  {Elimelech}},\ }\href@noop {} {\bibfield  {journal} {\bibinfo  {journal}
  {Journal of Membrane Science}\ }\textbf {\bibinfo {volume} {296}},\ \bibinfo
  {pages} {83} (\bibinfo {year} {2007})}\BibitemShut {NoStop}%
\bibitem [{\citenamefont {Linkhorst}\ \emph {et~al.}(2016)\citenamefont
  {Linkhorst}, \citenamefont {Beckmann}, \citenamefont {Go}, \citenamefont
  {Kuehne},\ and\ \citenamefont {Wessling}}]{linkhorst}%
  \BibitemOpen
  \bibfield  {author} {\bibinfo {author} {\bibfnamefont {J.}~\bibnamefont
  {Linkhorst}}, \bibinfo {author} {\bibfnamefont {T.}~\bibnamefont {Beckmann}},
  \bibinfo {author} {\bibfnamefont {D.}~\bibnamefont {Go}}, \bibinfo {author}
  {\bibfnamefont {A.~J.}\ \bibnamefont {Kuehne}}, \ and\ \bibinfo {author}
  {\bibfnamefont {M.}~\bibnamefont {Wessling}},\ }\href@noop {} {\bibfield
  {journal} {\bibinfo  {journal} {Scientific reports}\ }\textbf {\bibinfo
  {volume} {6}},\ \bibinfo {pages} {22376} (\bibinfo {year}
  {2016})}\BibitemShut {NoStop}%
\bibitem [{\citenamefont {Higgins}\ \emph {et~al.}(2007)\citenamefont
  {Higgins}, \citenamefont {Eddington}, \citenamefont {Bhatia},\ and\
  \citenamefont {Mahadevan}}]{higgins}%
  \BibitemOpen
  \bibfield  {author} {\bibinfo {author} {\bibfnamefont {J.}~\bibnamefont
  {Higgins}}, \bibinfo {author} {\bibfnamefont {D.}~\bibnamefont {Eddington}},
  \bibinfo {author} {\bibfnamefont {S.}~\bibnamefont {Bhatia}}, \ and\ \bibinfo
  {author} {\bibfnamefont {L.}~\bibnamefont {Mahadevan}},\ }\href@noop {}
  {\bibfield  {journal} {\bibinfo  {journal} {Proceedings of the National
  Academy of Sciences}\ }\textbf {\bibinfo {volume} {104}},\ \bibinfo {pages}
  {20496} (\bibinfo {year} {2007})}\BibitemShut {NoStop}%
\bibitem [{\citenamefont {Genovese}\ and\ \citenamefont
  {Sprakel}(2011)}]{genovese}%
  \BibitemOpen
  \bibfield  {author} {\bibinfo {author} {\bibfnamefont {D.}~\bibnamefont
  {Genovese}}\ and\ \bibinfo {author} {\bibfnamefont {J.}~\bibnamefont
  {Sprakel}},\ }\href@noop {} {\bibfield  {journal} {\bibinfo  {journal} {Soft
  Matter}\ }\textbf {\bibinfo {volume} {7}},\ \bibinfo {pages} {3889} (\bibinfo
  {year} {2011})}\BibitemShut {NoStop}%
\bibitem [{\citenamefont {Sauret}\ \emph {et~al.}(2018)\citenamefont {Sauret},
  \citenamefont {Somszor}, \citenamefont {Villermaux},\ and\ \citenamefont
  {Dressaire}}]{wyss}%
  \BibitemOpen
  \bibfield  {author} {\bibinfo {author} {\bibfnamefont {A.}~\bibnamefont
  {Sauret}}, \bibinfo {author} {\bibfnamefont {K.}~\bibnamefont {Somszor}},
  \bibinfo {author} {\bibfnamefont {E.}~\bibnamefont {Villermaux}}, \ and\
  \bibinfo {author} {\bibfnamefont {E.}~\bibnamefont {Dressaire}},\ }\href@noop
  {} {\bibfield  {journal} {\bibinfo  {journal} {Physical Review Fluids}\
  }\textbf {\bibinfo {volume} {3}},\ \bibinfo {pages} {104301} (\bibinfo {year}
  {2018})}\BibitemShut {NoStop}%
\bibitem [{\citenamefont {MacDonald}\ \emph {et~al.}(2003)\citenamefont
  {MacDonald}, \citenamefont {Spalding},\ and\ \citenamefont
  {Dholakia}}]{macdonald}%
  \BibitemOpen
  \bibfield  {author} {\bibinfo {author} {\bibfnamefont {M.~P.}\ \bibnamefont
  {MacDonald}}, \bibinfo {author} {\bibfnamefont {G.~C.}\ \bibnamefont
  {Spalding}}, \ and\ \bibinfo {author} {\bibfnamefont {K.}~\bibnamefont
  {Dholakia}},\ }\href@noop {} {\bibfield  {journal} {\bibinfo  {journal}
  {Nature}\ }\textbf {\bibinfo {volume} {426}},\ \bibinfo {pages} {421}
  (\bibinfo {year} {2003})}\BibitemShut {NoStop}%
\bibitem [{\citenamefont {Falk}\ \emph
  {et~al.}(2015{\natexlab{a}})\citenamefont {Falk}, \citenamefont {Pellenq},
  \citenamefont {Ulm},\ and\ \citenamefont {Coasne}}]{falk1}%
  \BibitemOpen
  \bibfield  {author} {\bibinfo {author} {\bibfnamefont {K.}~\bibnamefont
  {Falk}}, \bibinfo {author} {\bibfnamefont {R.}~\bibnamefont {Pellenq}},
  \bibinfo {author} {\bibfnamefont {F.~J.}\ \bibnamefont {Ulm}}, \ and\
  \bibinfo {author} {\bibfnamefont {B.}~\bibnamefont {Coasne}},\ }\href@noop {}
  {\bibfield  {journal} {\bibinfo  {journal} {Energy \& Fuels}\ }\textbf
  {\bibinfo {volume} {29}},\ \bibinfo {pages} {7889} (\bibinfo {year}
  {2015}{\natexlab{a}})}\BibitemShut {NoStop}%
\bibitem [{\citenamefont {Berthonneau}\ \emph {et~al.}(2018)\citenamefont
  {Berthonneau}, \citenamefont {Obliger}, \citenamefont {Valdenaire},
  \citenamefont {Grauby}, \citenamefont {Ferry}, \citenamefont {Chaudanson},
  \citenamefont {Levitz}, \citenamefont {Kim}, \citenamefont {Ulm},\ and\
  \citenamefont {Pellenq}}]{berthonneau}%
  \BibitemOpen
  \bibfield  {author} {\bibinfo {author} {\bibfnamefont {J.}~\bibnamefont
  {Berthonneau}}, \bibinfo {author} {\bibfnamefont {A.}~\bibnamefont
  {Obliger}}, \bibinfo {author} {\bibfnamefont {P.-L.}\ \bibnamefont
  {Valdenaire}}, \bibinfo {author} {\bibfnamefont {O.}~\bibnamefont {Grauby}},
  \bibinfo {author} {\bibfnamefont {D.}~\bibnamefont {Ferry}}, \bibinfo
  {author} {\bibfnamefont {D.}~\bibnamefont {Chaudanson}}, \bibinfo {author}
  {\bibfnamefont {P.}~\bibnamefont {Levitz}}, \bibinfo {author} {\bibfnamefont
  {J.~J.}\ \bibnamefont {Kim}}, \bibinfo {author} {\bibfnamefont {F.-J.}\
  \bibnamefont {Ulm}}, \ and\ \bibinfo {author} {\bibfnamefont {R.~J.-M.}\
  \bibnamefont {Pellenq}},\ }\href@noop {} {\bibfield  {journal} {\bibinfo
  {journal} {Proceedings of the National Academy of Sciences}\ }\textbf
  {\bibinfo {volume} {115}},\ \bibinfo {pages} {12365} (\bibinfo {year}
  {2018})}\BibitemShut {NoStop}%
\bibitem [{\citenamefont {Lin}\ \emph {et~al.}(2016)\citenamefont {Lin},
  \citenamefont {Liu}, \citenamefont {Liang}, \citenamefont {Lee},
  \citenamefont {Sun}, \citenamefont {Wang}, \citenamefont {Yan}, \citenamefont
  {Xie},\ and\ \citenamefont {Cui}}]{lin}%
  \BibitemOpen
  \bibfield  {author} {\bibinfo {author} {\bibfnamefont {D.}~\bibnamefont
  {Lin}}, \bibinfo {author} {\bibfnamefont {Y.}~\bibnamefont {Liu}}, \bibinfo
  {author} {\bibfnamefont {Z.}~\bibnamefont {Liang}}, \bibinfo {author}
  {\bibfnamefont {H.-W.}\ \bibnamefont {Lee}}, \bibinfo {author} {\bibfnamefont
  {J.}~\bibnamefont {Sun}}, \bibinfo {author} {\bibfnamefont {H.}~\bibnamefont
  {Wang}}, \bibinfo {author} {\bibfnamefont {K.}~\bibnamefont {Yan}}, \bibinfo
  {author} {\bibfnamefont {J.}~\bibnamefont {Xie}}, \ and\ \bibinfo {author}
  {\bibfnamefont {Y.}~\bibnamefont {Cui}},\ }\href@noop {} {\bibfield
  {journal} {\bibinfo  {journal} {Nature nanotechnology}\ }\textbf {\bibinfo
  {volume} {11}},\ \bibinfo {pages} {626} (\bibinfo {year} {2016})}\BibitemShut
  {NoStop}%
\bibitem [{\citenamefont {Borchardt}\ \emph {et~al.}(2018)\citenamefont
  {Borchardt}, \citenamefont {Leistenschneider}, \citenamefont {Haase},\ and\
  \citenamefont {Dvoyashkin}}]{borchardt}%
  \BibitemOpen
  \bibfield  {author} {\bibinfo {author} {\bibfnamefont {L.}~\bibnamefont
  {Borchardt}}, \bibinfo {author} {\bibfnamefont {D.}~\bibnamefont
  {Leistenschneider}}, \bibinfo {author} {\bibfnamefont {J.}~\bibnamefont
  {Haase}}, \ and\ \bibinfo {author} {\bibfnamefont {M.}~\bibnamefont
  {Dvoyashkin}},\ }\href@noop {} {\bibfield  {journal} {\bibinfo  {journal}
  {Advanced Energy Materials}\ }\textbf {\bibinfo {volume} {8}},\ \bibinfo
  {pages} {1800892} (\bibinfo {year} {2018})}\BibitemShut {NoStop}%
\bibitem [{\citenamefont {Mitchell}\ \emph {et~al.}(2015)\citenamefont
  {Mitchell}, \citenamefont {Pinar}, \citenamefont {Kenvin}, \citenamefont
  {Crivelli}, \citenamefont {K{\"a}rger},\ and\ \citenamefont
  {P{\'e}rez-Ram{\'\i}rez}}]{mitchell}%
  \BibitemOpen
  \bibfield  {author} {\bibinfo {author} {\bibfnamefont {S.}~\bibnamefont
  {Mitchell}}, \bibinfo {author} {\bibfnamefont {A.~B.}\ \bibnamefont {Pinar}},
  \bibinfo {author} {\bibfnamefont {J.}~\bibnamefont {Kenvin}}, \bibinfo
  {author} {\bibfnamefont {P.}~\bibnamefont {Crivelli}}, \bibinfo {author}
  {\bibfnamefont {J.}~\bibnamefont {K{\"a}rger}}, \ and\ \bibinfo {author}
  {\bibfnamefont {J.}~\bibnamefont {P{\'e}rez-Ram{\'\i}rez}},\ }\href@noop {}
  {\bibfield  {journal} {\bibinfo  {journal} {Nature communications}\ }\textbf
  {\bibinfo {volume} {6}},\ \bibinfo {pages} {1} (\bibinfo {year}
  {2015})}\BibitemShut {NoStop}%
\bibitem [{\citenamefont {Chanut}\ \emph {et~al.}(2016)\citenamefont {Chanut},
  \citenamefont {Wiersum}, \citenamefont {Lee}, \citenamefont {Hwang},
  \citenamefont {Ragon}, \citenamefont {Chevreau}, \citenamefont {Bourrelly},
  \citenamefont {Kuchta}, \citenamefont {Chang}, \citenamefont {Serre} \emph
  {et~al.}}]{chanut}%
  \BibitemOpen
  \bibfield  {author} {\bibinfo {author} {\bibfnamefont {N.}~\bibnamefont
  {Chanut}}, \bibinfo {author} {\bibfnamefont {A.~D.}\ \bibnamefont {Wiersum}},
  \bibinfo {author} {\bibfnamefont {U.-H.}\ \bibnamefont {Lee}}, \bibinfo
  {author} {\bibfnamefont {Y.~K.}\ \bibnamefont {Hwang}}, \bibinfo {author}
  {\bibfnamefont {F.}~\bibnamefont {Ragon}}, \bibinfo {author} {\bibfnamefont
  {H.}~\bibnamefont {Chevreau}}, \bibinfo {author} {\bibfnamefont
  {S.}~\bibnamefont {Bourrelly}}, \bibinfo {author} {\bibfnamefont
  {B.}~\bibnamefont {Kuchta}}, \bibinfo {author} {\bibfnamefont {J.-S.}\
  \bibnamefont {Chang}}, \bibinfo {author} {\bibfnamefont {C.}~\bibnamefont
  {Serre}},  \emph {et~al.},\ }\href@noop {} {\bibfield  {journal} {\bibinfo
  {journal} {European Journal of Inorganic Chemistry}\ }\textbf {\bibinfo
  {volume} {2016}},\ \bibinfo {pages} {4416} (\bibinfo {year}
  {2016})}\BibitemShut {NoStop}%
\bibitem [{\citenamefont {Xu}\ \emph {et~al.}(2020)\citenamefont {Xu},
  \citenamefont {Gao}, \citenamefont {Chen},\ and\ \citenamefont {Yan}}]{zxu}%
  \BibitemOpen
  \bibfield  {author} {\bibinfo {author} {\bibfnamefont {Z.}~\bibnamefont
  {Xu}}, \bibinfo {author} {\bibfnamefont {L.}~\bibnamefont {Gao}}, \bibinfo
  {author} {\bibfnamefont {P.}~\bibnamefont {Chen}}, \ and\ \bibinfo {author}
  {\bibfnamefont {L.-T.}\ \bibnamefont {Yan}},\ }\href@noop {} {\bibfield
  {journal} {\bibinfo  {journal} {Soft Matter}\ }\textbf {\bibinfo {volume}
  {16}},\ \bibinfo {pages} {3869} (\bibinfo {year} {2020})}\BibitemShut
  {NoStop}%
\bibitem [{\citenamefont {Rouquerol}\ \emph {et~al.}(1994)\citenamefont
  {Rouquerol}, \citenamefont {Avnir}, \citenamefont {Fairbridge}, \citenamefont
  {Everett}, \citenamefont {Haynes}, \citenamefont {Pernicone}, \citenamefont
  {Ramsay}, \citenamefont {Sing},\ and\ \citenamefont {Unger}}]{rouquerol}%
  \BibitemOpen
  \bibfield  {author} {\bibinfo {author} {\bibfnamefont {J.}~\bibnamefont
  {Rouquerol}}, \bibinfo {author} {\bibfnamefont {D.}~\bibnamefont {Avnir}},
  \bibinfo {author} {\bibfnamefont {C.}~\bibnamefont {Fairbridge}}, \bibinfo
  {author} {\bibfnamefont {D.}~\bibnamefont {Everett}}, \bibinfo {author}
  {\bibfnamefont {J.}~\bibnamefont {Haynes}}, \bibinfo {author} {\bibfnamefont
  {N.}~\bibnamefont {Pernicone}}, \bibinfo {author} {\bibfnamefont
  {J.}~\bibnamefont {Ramsay}}, \bibinfo {author} {\bibfnamefont
  {K.}~\bibnamefont {Sing}}, \ and\ \bibinfo {author} {\bibfnamefont
  {K.}~\bibnamefont {Unger}},\ }\href@noop {} {\bibfield  {journal} {\bibinfo
  {journal} {Pure and Applied Chemistry}\ }\textbf {\bibinfo {volume} {66}},\
  \bibinfo {pages} {1739} (\bibinfo {year} {1994})}\BibitemShut {NoStop}%
\bibitem [{\citenamefont {Frentrup}\ \emph {et~al.}(2012)\citenamefont
  {Frentrup}, \citenamefont {Avenda{\~n}o}, \citenamefont {Horsch},
  \citenamefont {Salih},\ and\ \citenamefont {M{\"u}ller}}]{frentrup}%
  \BibitemOpen
  \bibfield  {author} {\bibinfo {author} {\bibfnamefont {H.}~\bibnamefont
  {Frentrup}}, \bibinfo {author} {\bibfnamefont {C.}~\bibnamefont
  {Avenda{\~n}o}}, \bibinfo {author} {\bibfnamefont {M.}~\bibnamefont
  {Horsch}}, \bibinfo {author} {\bibfnamefont {A.}~\bibnamefont {Salih}}, \
  and\ \bibinfo {author} {\bibfnamefont {E.~A.}\ \bibnamefont {M{\"u}ller}},\
  }\href@noop {} {\bibfield  {journal} {\bibinfo  {journal} {Molecular
  Simulation}\ }\textbf {\bibinfo {volume} {38}},\ \bibinfo {pages} {540}
  (\bibinfo {year} {2012})}\BibitemShut {NoStop}%
\bibitem [{\citenamefont {Sholl}(1999)}]{sholl}%
  \BibitemOpen
  \bibfield  {author} {\bibinfo {author} {\bibfnamefont {D.~S.}\ \bibnamefont
  {Sholl}},\ }\href@noop {} {\bibfield  {journal} {\bibinfo  {journal}
  {Chemical Engineering Journal}\ }\textbf {\bibinfo {volume} {74}},\ \bibinfo
  {pages} {25} (\bibinfo {year} {1999})}\BibitemShut {NoStop}%
\bibitem [{\citenamefont {Valiullin}\ \emph {et~al.}(2006)\citenamefont
  {Valiullin}, \citenamefont {Naumov}, \citenamefont {Galvosas}, \citenamefont
  {K{\"a}rger}, \citenamefont {Woo}, \citenamefont {Porcheron},\ and\
  \citenamefont {Monson}}]{valuilin}%
  \BibitemOpen
  \bibfield  {author} {\bibinfo {author} {\bibfnamefont {R.}~\bibnamefont
  {Valiullin}}, \bibinfo {author} {\bibfnamefont {S.}~\bibnamefont {Naumov}},
  \bibinfo {author} {\bibfnamefont {P.}~\bibnamefont {Galvosas}}, \bibinfo
  {author} {\bibfnamefont {J.}~\bibnamefont {K{\"a}rger}}, \bibinfo {author}
  {\bibfnamefont {H.-J.}\ \bibnamefont {Woo}}, \bibinfo {author} {\bibfnamefont
  {F.}~\bibnamefont {Porcheron}}, \ and\ \bibinfo {author} {\bibfnamefont
  {P.~A.}\ \bibnamefont {Monson}},\ }\href@noop {} {\bibfield  {journal}
  {\bibinfo  {journal} {Nature}\ }\textbf {\bibinfo {volume} {443}},\ \bibinfo
  {pages} {965} (\bibinfo {year} {2006})}\BibitemShut {NoStop}%
\bibitem [{\citenamefont {Falk}\ \emph
  {et~al.}(2015{\natexlab{b}})\citenamefont {Falk}, \citenamefont {Coasne},
  \citenamefont {Pellenq}, \citenamefont {Ulm},\ and\ \citenamefont
  {Bocquet}}]{falk2_ar}%
  \BibitemOpen
  \bibfield  {author} {\bibinfo {author} {\bibfnamefont {K.}~\bibnamefont
  {Falk}}, \bibinfo {author} {\bibfnamefont {B.}~\bibnamefont {Coasne}},
  \bibinfo {author} {\bibfnamefont {R.}~\bibnamefont {Pellenq}}, \bibinfo
  {author} {\bibfnamefont {F.-J.}\ \bibnamefont {Ulm}}, \ and\ \bibinfo
  {author} {\bibfnamefont {L.}~\bibnamefont {Bocquet}},\ }\href@noop {}
  {\bibfield  {journal} {\bibinfo  {journal} {Nature communications}\ }\textbf
  {\bibinfo {volume} {6}},\ \bibinfo {pages} {1} (\bibinfo {year}
  {2015}{\natexlab{b}})}\BibitemShut {NoStop}%
\bibitem [{\citenamefont {Travis}\ \emph {et~al.}(1997)\citenamefont {Travis},
  \citenamefont {Todd},\ and\ \citenamefont {Evans}}]{travis}%
  \BibitemOpen
  \bibfield  {author} {\bibinfo {author} {\bibfnamefont {K.~P.}\ \bibnamefont
  {Travis}}, \bibinfo {author} {\bibfnamefont {B.~D.}~\bibnamefont {Todd}}, \ and\
  \bibinfo {author} {\bibfnamefont {D.~J.}\ \bibnamefont {Evans}},\ }\href@noop
  {} {\bibfield  {journal} {\bibinfo  {journal} {Physical Review E}\ }\textbf
  {\bibinfo {volume} {55}},\ \bibinfo {pages} {4288} (\bibinfo {year}
  {1997})}\BibitemShut {NoStop}%
\bibitem [{\citenamefont {Bhatia}\ \emph {et~al.}(2011)\citenamefont {Bhatia},
  \citenamefont {Bonilla},\ and\ \citenamefont {Nicholson}}]{bhatia}%
  \BibitemOpen
  \bibfield  {author} {\bibinfo {author} {\bibfnamefont {S.~K.}\ \bibnamefont
  {Bhatia}}, \bibinfo {author} {\bibfnamefont {M.~R.}\ \bibnamefont {Bonilla}},
  \ and\ \bibinfo {author} {\bibfnamefont {D.}~\bibnamefont {Nicholson}},\
  }\href@noop {} {\bibfield  {journal} {\bibinfo  {journal} {Physical Chemistry
  Chemical Physics}\ }\textbf {\bibinfo {volume} {13}},\ \bibinfo {pages}
  {15350} (\bibinfo {year} {2011})}\BibitemShut {NoStop}%
\bibitem [{\citenamefont {Aguirre}\ \emph {et~al.}(2010)\citenamefont
  {Aguirre}, \citenamefont {Grande}, \citenamefont {Calvo}, \citenamefont
  {Pugnaloni},\ and\ \citenamefont {G{\'e}minard}}]{aguirre}%
  \BibitemOpen
  \bibfield  {author} {\bibinfo {author} {\bibfnamefont {M.~A.}\ \bibnamefont
  {Aguirre}}, \bibinfo {author} {\bibfnamefont {J.~G.}\ \bibnamefont {Grande}},
  \bibinfo {author} {\bibfnamefont {A.}~\bibnamefont {Calvo}}, \bibinfo
  {author} {\bibfnamefont {L.~A.}\ \bibnamefont {Pugnaloni}}, \ and\ \bibinfo
  {author} {\bibfnamefont {J.-C.}\ \bibnamefont {G{\'e}minard}},\ }\href@noop
  {} {\bibfield  {journal} {\bibinfo  {journal} {Physical review letters}\
  }\textbf {\bibinfo {volume} {104}},\ \bibinfo {pages} {238002} (\bibinfo
  {year} {2010})}\BibitemShut {NoStop}%
\bibitem [{\citenamefont {Beverloo}\ \emph {et~al.}(1961)\citenamefont
  {Beverloo}, \citenamefont {Leniger},\ and\ \citenamefont {Van~de
  Velde}}]{beverloo}%
  \BibitemOpen
  \bibfield  {author} {\bibinfo {author} {\bibfnamefont {W.~A.}\ \bibnamefont
  {Beverloo}}, \bibinfo {author} {\bibfnamefont {H.~A.}\ \bibnamefont
  {Leniger}}, \ and\ \bibinfo {author} {\bibfnamefont {J.}~\bibnamefont {Van~de
  Velde}},\ }\href@noop {} {\bibfield  {journal} {\bibinfo  {journal} {Chemical
  engineering science}\ }\textbf {\bibinfo {volume} {15}},\ \bibinfo {pages}
  {260} (\bibinfo {year} {1961})}\BibitemShut {NoStop}%
\bibitem [{\citenamefont {Nedderman}\ \emph {et~al.}(1982)\citenamefont
  {Nedderman}, \citenamefont {T{\"u}z{\"u}n}, \citenamefont {Savage},\ and\
  \citenamefont {Houlsby}}]{nedderman}%
  \BibitemOpen
  \bibfield  {author} {\bibinfo {author} {\bibfnamefont {R.}~\bibnamefont
  {Nedderman}}, \bibinfo {author} {\bibfnamefont {U.}~\bibnamefont
  {T{\"u}z{\"u}n}}, \bibinfo {author} {\bibfnamefont {S.}~\bibnamefont
  {Savage}}, \ and\ \bibinfo {author} {\bibfnamefont {G.}~\bibnamefont
  {Houlsby}},\ }\href@noop {} {\bibfield  {journal} {\bibinfo  {journal} {Chem.
  Eng. Sci}\ }\textbf {\bibinfo {volume} {37}},\ \bibinfo {pages} {1597}
  (\bibinfo {year} {1982})}\BibitemShut {NoStop}%
\bibitem [{\citenamefont {Hagen}(1852)}]{hagen}%
  \BibitemOpen
  \bibfield  {author} {\bibinfo {author} {\bibfnamefont {G.}~\bibnamefont
  {Hagen}},\ }\href@noop {} {\bibfield  {journal} {\bibinfo  {journal}
  {Preussischen Akademie der Wissenschaften zu Berlin}\ }\textbf {\bibinfo
  {volume} {35}},\ \bibinfo {pages} {35} (\bibinfo {year} {1852})}\BibitemShut
  {NoStop}%
\bibitem [{\citenamefont {Wu}\ \emph {et~al.}(2016)\citenamefont {Wu},
  \citenamefont {Chen}, \citenamefont {Li},\ and\ \citenamefont {Dong}}]{wu}%
  \BibitemOpen
  \bibfield  {author} {\bibinfo {author} {\bibfnamefont {K.}~\bibnamefont
  {Wu}}, \bibinfo {author} {\bibfnamefont {Z.}~\bibnamefont {Chen}}, \bibinfo
  {author} {\bibfnamefont {X.}~\bibnamefont {Li}}, \ and\ \bibinfo {author}
  {\bibfnamefont {X.}~\bibnamefont {Dong}},\ }\href@noop {} {\bibfield
  {journal} {\bibinfo  {journal} {Scientific reports}\ }\textbf {\bibinfo
  {volume} {6}},\ \bibinfo {pages} {33461} (\bibinfo {year}
  {2016})}\BibitemShut {NoStop}%
\bibitem [{\citenamefont {Shen}\ \emph {et~al.}(2019)\citenamefont {Shen},
  \citenamefont {Song}, \citenamefont {Hu}, \citenamefont {Zhu},\ and\
  \citenamefont {Zhu}}]{shen}%
  \BibitemOpen
  \bibfield  {author} {\bibinfo {author} {\bibfnamefont {W.}~\bibnamefont
  {Shen}}, \bibinfo {author} {\bibfnamefont {F.}~\bibnamefont {Song}}, \bibinfo
  {author} {\bibfnamefont {X.}~\bibnamefont {Hu}}, \bibinfo {author}
  {\bibfnamefont {G.}~\bibnamefont {Zhu}}, \ and\ \bibinfo {author}
  {\bibfnamefont {W.}~\bibnamefont {Zhu}},\ }\href@noop {} {\bibfield
  {journal} {\bibinfo  {journal} {Scientific reports}\ }\textbf {\bibinfo
  {volume} {9}},\ \bibinfo {pages} {1} (\bibinfo {year} {2019})}\BibitemShut
  {NoStop}%
\end{thebibliography}
\providecommand{\noopsort}[1]{}\providecommand{\singleletter}[1]{#1}%
\end{document}


\preprint{APS/123-QED}

\title{Supplemental Materials for "Granular-like behavior of molecular flow in constricted nanopores"}

\author{Y. Magnin}
\email{magnin@mit.edu}
\affiliation{MIT, Energy Initiative, Massachusetts Institute of Technology, Cambridge, MA, United States}
\author{N. Chanut}
\affiliation{MIT, Energy Initiative, Massachusetts Institute of Technology, Cambridge, MA, United States}
\author{D.A. Pantano}
\affiliation{Total Marketing \& Services, Chemin du Canal, 69360 Solaize, France}
\author{E. Chaput}
\affiliation{Total E\&P Research \& D\'evelopement, CSTJF - Avenue Larribau, 64018 Pau Cedex, France}
\author{F-J. Ulm}
\affiliation{MIT, Energy Initiative, Massachusetts Institute of Technology, Cambridge, MA, United States}
\author{R. J.-M. Pellenq}
\affiliation{Multi-Scale Materials Science for Energy and Environment the MIT / CNRS / Aix-Marseille University Joint Laboratory at Massachusetts Institute of Technology, Cambridge, Massachusetts 02139, USA}
\altaffiliation[Also at ]{Department of Physics, Georgetown University, Washington DC, USA}
\author{E. Villermaux}
\affiliation{Aix Marseille Universit\'e, CNRS, Centrale Marseille, IRPHE UMR 7342, Marseille 13384, France}
\date{\today}
\maketitle

\section*{Methods}
\begin{figure}
\begin{center}
\includegraphics[scale=0.4]{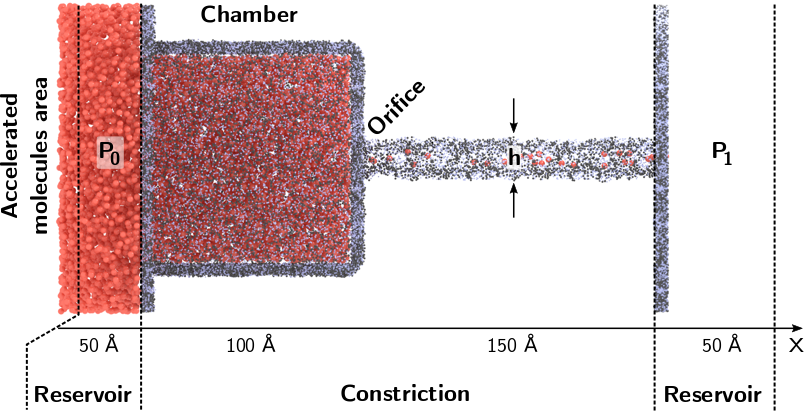}
\end{center}
\caption{\label{sifig0} Atomistic structures of constriction designed to simulate the molecular flow, and made of carbon (dark) and hydrogen atoms (blue).}
\end{figure}

Synthetic porous structures are composed by a chamber of 100 \AA\ edges, connected by one or few pores with diameters $h$ ranging from 10 \AA\ to 50\AA. Such systems are shaped in a dense amorphous and nanoporous carbon unit-cell of about 50 \AA\ edges. Unit-cell are initially generated by the Hybrid Reverse Monte Carlo technique described elsewhere \cite{Opletal,Jain,Bousige}. In this approach, the initial configuration is prepared from a random distribution of atoms in the simulation box. The next step consists in minimizing the configurational energy and reached the thermodynamical equilibrium, as in a standard Monte Carlo simulation. In addition, the convergence is also constrained by experimental data as the radial distribution function, obtained by the inverse Fourier transform of the structure factor S(Q), measured by neutron scattering experiments. Theoretically, this inverse method allows to reconstruct a realistic atomistic structures of experimental samples. The carbon unit-cell is then duplicated in each direction to generate a 150$\times$150$\times$300 \AA$^3$\ amorphous solid. The chamber and the pore are then shaped into the structure by removing carbon and hydrogen atoms to produce the meso-constrictions.\\
Methane adsorption is performed by Monte Carlo simulations in the Grand Canonical ensemble \cite{adams}. In this algorithm, the porous adsorbent randomly exchanges molecules with an ideal gas reservoir at constant chemical potential ($\mu_{CH_4}$), volume ($V$) and temperature ($T$). The system is first equilibrated during $1.10^6$ cycles. A cycle consists in trying 1000 molecules insertions or 1000 molecules deletions. Such a choice is determined randomly from a normal distribution with a 50\% insertion or deletion probability at each cycle. During equilibration, molecules are adsorbed into the left reservoir, the chamber and the pores. When the equilibrium is reached, the density of molecules is directly related to the initial set of thermodynamical parameters in the Grand Canonical ensemble ($\mu_{CH_4}$,V,T). Typically, calculations presented in this work correspond to an adsorption at a pressure around the MPa and the temperature is fixed to 400K for each simulation.\\
Both side of the structure is connected to a reservoir, with a difference in density ensuring a pressure difference ($\Delta P$) between the two reservoirs and inducing the molecular flow through the system. In order to maintain the flow, a constant force is imposed to molecules between $x=0$ and $x=10$ \AA\ in the left reservoir, Fig.\ref{sifig0}. Elsewhere, the force is set to zero and the accelerated molecules are accumulated in the chamber downstream \cite{frentrup}. The porous matrix is described within the frame of the frozen matrix approximation \textit{i.e.} the carbon and hydrogen atoms of the pore walls are not subjected to the thermal agitation. Methane molecules interact with their surrounding neighbors (carbon, hydrogen atoms and other methane molecules) with a long range Lennard-Jones potential. 

\begin{eqnarray}
U = 4\ \epsilon_{ij} \left[ \left(\frac{\sigma_{ij}}{r_{ij}}\right)^{12} - \left(\frac{\sigma_{ij}}{r_{ij}}\right)^{6}\right]
\end{eqnarray}
The Lennard-Jones parameters for the different atomic species, Table \ref{tbl:lj}, are determined by the Lorentz-Berthelot mixing rule as, 

\begin{eqnarray}
\sigma_{ij}=\left(\frac{\sigma_i\ +\ \sigma_j}{2}\right)
\end{eqnarray}
and

\begin{eqnarray}
\epsilon_{ij}=\sqrt{\epsilon_i \epsilon_j},
\end{eqnarray}
with the following parameters:

\begin{table}[h]
\begin{center}
\small
  \caption{\ Lennard-Jones parameters for carbon, hydrogen and methane molecules.}
  \label{tbl:lj}
  \begin{tabular*}{0.48\textwidth}{@{\extracolsep{\fill}}|lll|}
    \hline
    Atoms and pseudo atoms \hspace{0.25cm} & $\sigma$ (\AA)\hspace{0.25cm} & $\epsilon$ (meV) \\
    \hline
    CH$_4$ & 3.73 & 12.7 \\
    C      & 3.36 & 2.4 \\
    H      & 2.42 & 1.3 \\
    \hline
  \end{tabular*}
\end{center}
\end{table}

\noindent During the out-of-equilibrium ($N$,$V$,$T$) molecular dynamics, the simulation step following the GCMC gas adsorption procedure, a Nos\'e-Hoover thermostat ensures a constant temperature set to 400$K$. Thus, the system dissipate energy under the form of heat, using a thermal inertia parameter, which determines the rate of the heat transfer to the outside, in such a way that the overall (averaged) system temperature is maintained constant during the flow simulations. This mimics the thermalization through collisions and adsorption of molecules to the systems walls (the porous matrix in our case). A full simulation consists in a run of 10ns. After a transient in the first 5ns, a steady state is reached, lasting 5 ns, where averaged quantities like molecular velocity $v$, pressure $P$ and fluid density $\rho$ are measured. Simulations are performed for different differential pressure $\Delta P$ (ranging from $\sim$ 4 to 24 MPa, around a mean pressure $P$ of the order of a MPa itself), pore diameters and connectivity $\xi$ (number of branching pores), in system with and without constrictions.

\section*{Transport relationship for compressible fluids}
\begin{figure}
\begin{center}
\includegraphics[scale=0.56]{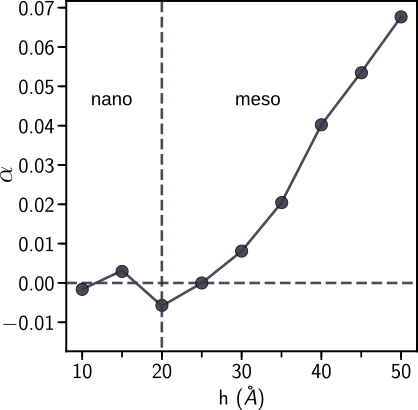}
\end{center}
\caption{\label{sifig8} Non-linearity correction exponent $\alpha$ as a function of the pore diameter.}
\end{figure}

\begin{figure}
\begin{center}
\includegraphics[scale=0.35]{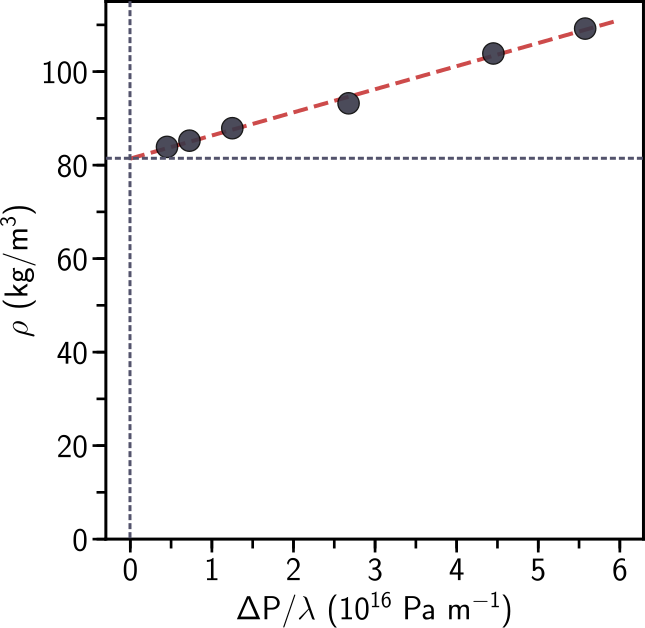}
\end{center}
\caption{\label{sifig2} Linear evolution of the fluid density (calculated in the entire system) as a function of the pressure gradient $\Delta P/\lambda$. The red dashed line corresponds to the linear fit of the simulation sresults (dark gray circles).}
\end{figure}

\begin{figure}
\begin{center}
\includegraphics[scale=0.5]{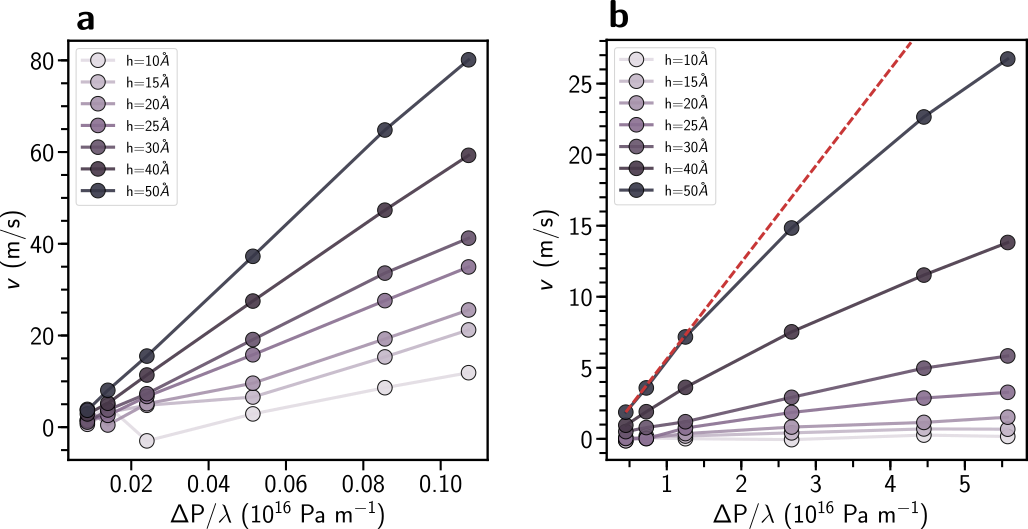}
\end{center}
\caption{\label{sifig3} a. Velocity of methane molecules as a function of the pressure gradient, for different pore diameters, in a system without constrictions. b. Same than (a) through constricted systems. The red dashed line indicates the linear part of $v$ at moderate pressure, and the slight linear deviation of $v$.}
\end{figure}

In order to propose an extended expression for the velocity $v$ and the mass flow rate $W$ for compressible fluids in constricted systems, we show that the density is a linear function of the pressure gradient, Fig.\ref{sifig2}. As can be seen, in a pressure ranging from 5 to 25 MPa (caracteristic in shale gas applications), the density scales linearly and is fitted as,
\begin{eqnarray}
\label{eq:4}
\rho= \zeta\,\Delta P/\lambda+ \rho_{\rm eq},
\end{eqnarray}
with $\zeta$ the fluid compressibility and $\rho_{\rm eq}$ the equilibrium density obtained at $\Delta P=0$.\\
In order to propose a transport relationship accounting for the fluid compressibility, we write the Darcy-like equation for a constricted system in a non-linear form, by  correcting the pressure gradient in $v$ with the exponent $\alpha$ as:
\begin{eqnarray}
\label{eq:5}
v\ =\ \beta \left(\frac{\Delta P}{\lambda}\right)^{1-\alpha}.
\end{eqnarray}
Thus, $\alpha$ allows to account for the non-linear behavior of $v$ shown in the Fig.\ref{sifig3}b. Coupling Beverloo's equation: 
\begin{eqnarray}
W\ &=&\ C\ \sqrt{\rho \Delta P/ \lambda}\ (h-k\sigma)^{5/2}\ =\ \rho v S,
\end{eqnarray}
with \eqref{eq:4} and \eqref{eq:5}, we write $C$ and $W$ in the case of compressible fluids as,
\begin{eqnarray}
\label{eq:Cdefcomp}
C\ &=&\ C_0\ S\ \sqrt{\left(\zeta\ \frac{\Delta P}{\lambda}\ +\ \rho_{eq}\right) \left(\frac{\Delta P}{\lambda}\right)^{1-2\alpha}},\\
W\ &=&\ C_0\ S\ (h-k\sigma)^{5/2}\ \left(\zeta\ \frac{\Delta P}{\lambda}\ +\ \rho_{eq}\right) \left(\frac{\Delta P}{\lambda}\right)^{1-\alpha}
\end{eqnarray}
The exponent $\alpha$ is found to solely depends on $h$, Fig.\ref{sifig8}. In simulations performed in this work, with a limited compressibility $\alpha \ll 1$, that we have first neglected for a sake of simplicity. However, extrapolating our result for compressible fluid, $\alpha$ should be accounted. In such a case, we can show that if $\zeta\ \frac{\Delta P}{\lambda}\ \gg\ \rho_{eq}$, $W$ should reads as,
\begin{eqnarray}
\label{eq:wlimh}
W\ \sim\ C_0\ S\ \zeta\ (h-k\sigma)^{5/2}\ \left(\frac{\Delta P}{\lambda}\right)^{1-\alpha}.
\end{eqnarray}

\begin{table}[h]
\small
\begin{center}
  \caption{Numerical values of the different parameters of the model.}
  \label{tbl:num}
  \begin{tabular*}{0.48\textwidth}{@{\extracolsep{\fill}}|lll|}
    \hline 
    $\zeta$ &(kg.m$^{-3}$Pa$^{-1}$) & 4.9.10$^{-16}$\\ 
    $\rho_{eq}$ &(kg.m$^{-3}$) & 81\\
    $k$ &(n.u) & 5.4 \\
    $C_0$ &(Pa$^{-1}$.s$^{-1}$.m$^{-1/2}$) & 7.1.10$^{6}$ \\
    \hline
  \end{tabular*}
  \end{center}
\end{table}

\begin{figure}
\begin{center}
\includegraphics[scale=0.3]{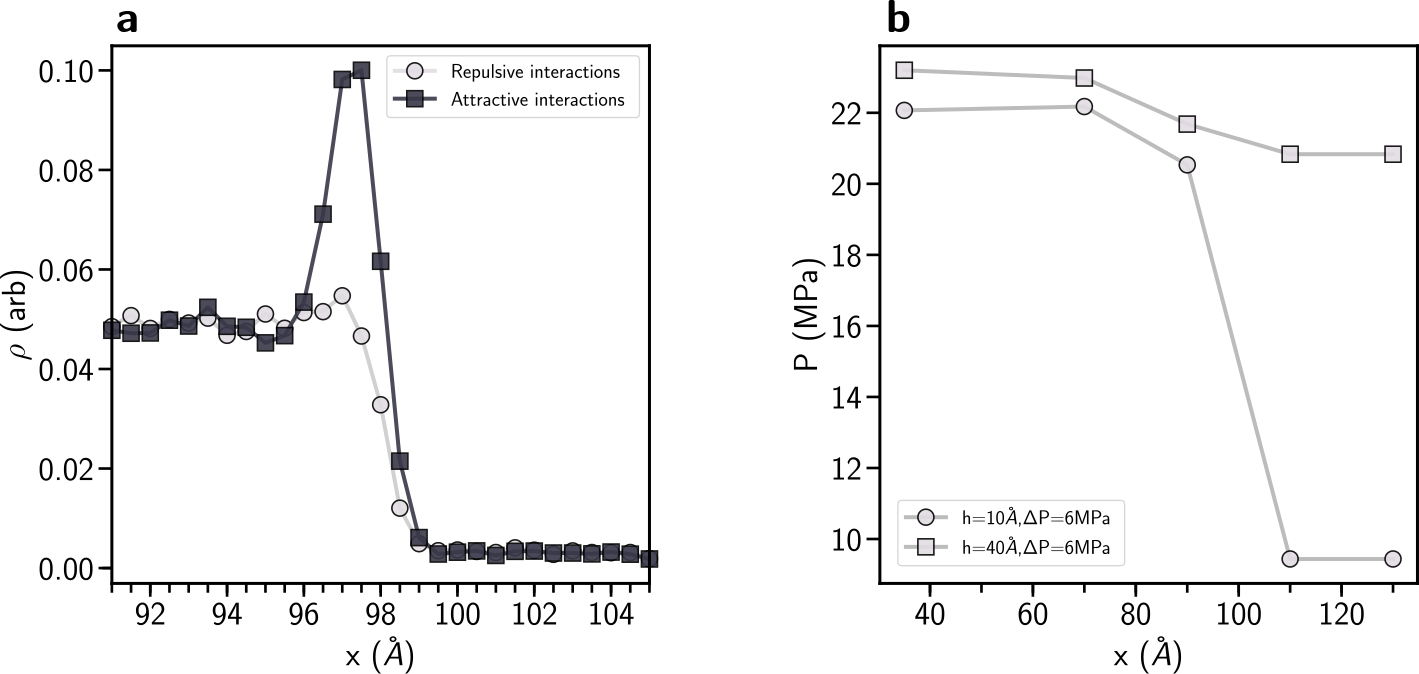}
\end{center}
\caption{\label{sifig1} Evolution of the fluid density and pressure along the flowing direction with attractive and repulsive interactions. a. Density of molecules along the direction $x$ using an attractive potential (dark gray squares) and a repulsive potential (light gray circles). b. Non-linear pressure decrease measured in constricted systems along the flowing direction $x$ with repulsive interactions. The circles correspond to a pore size $h$=10\AA, while squares correspond to $h$=40\AA\ with pressure difference $\Delta P$=6 MPa.}
\end{figure}

\begin{figure}
\begin{center}
\includegraphics[scale=0.32]{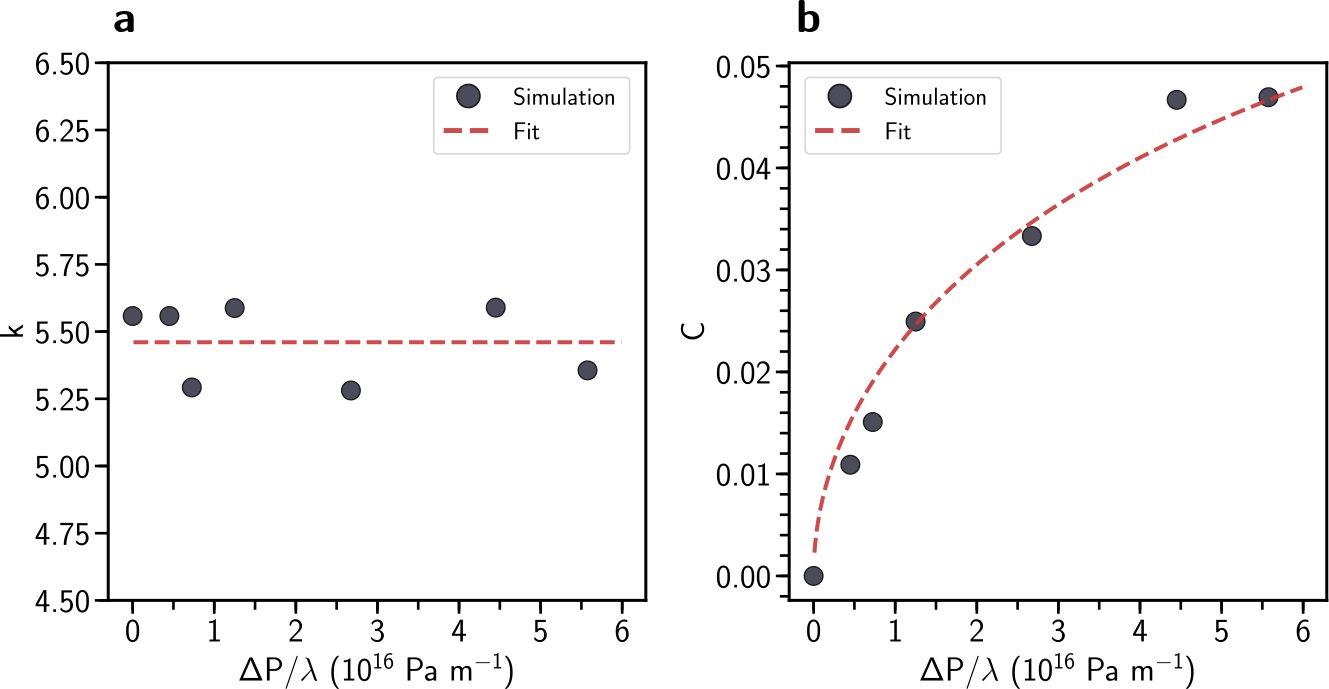}
\end{center}
\caption{\label{sifig4} Fit of Beverloo's parameters as a function of the pressure gradient. a. Evolution of the Beverloo's parameter $k$ as a function of the pressure gradient from simulations (dark gray circles). $k$ is found to be constant. b. Evolution of the Beverloo's parameter $C$ as a function of the pressure gradient from simulations (dark gray circles). This parameter is found to be $\propto \sqrt{\Delta P/\lambda}$. In figures (a) and (b), the red dashed lines correspond to the fit of the simulation results.}
\end{figure}

\begin{figure}
\begin{center}
\includegraphics[scale=0.4]{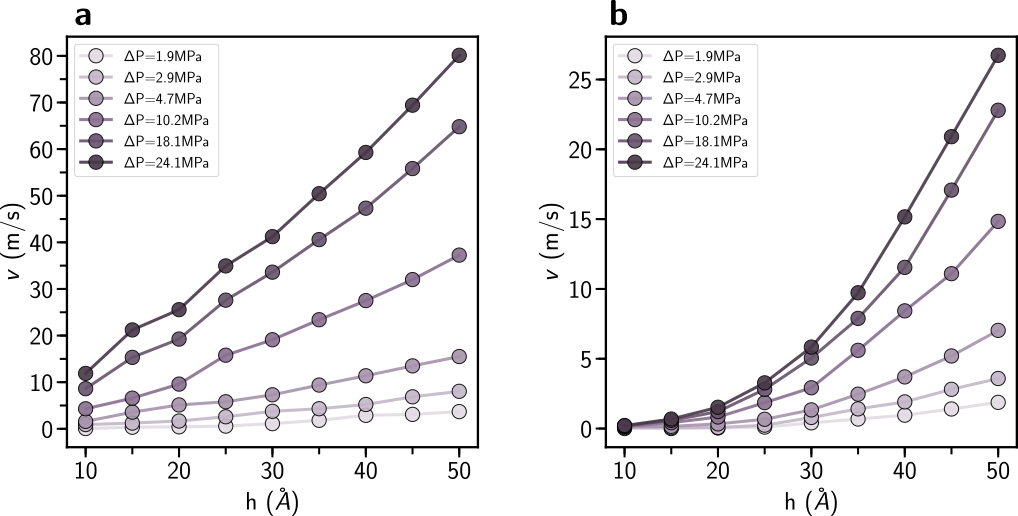}
\end{center}
\caption{\label{sifig5} a. Velocity of methane molecules as a function of pore diameters, for different pressure gradients in a system without constrictions. b. Same than (a) through constrictions.}
\end{figure}

\begin{figure}
\begin{center}
\includegraphics[scale=0.4]{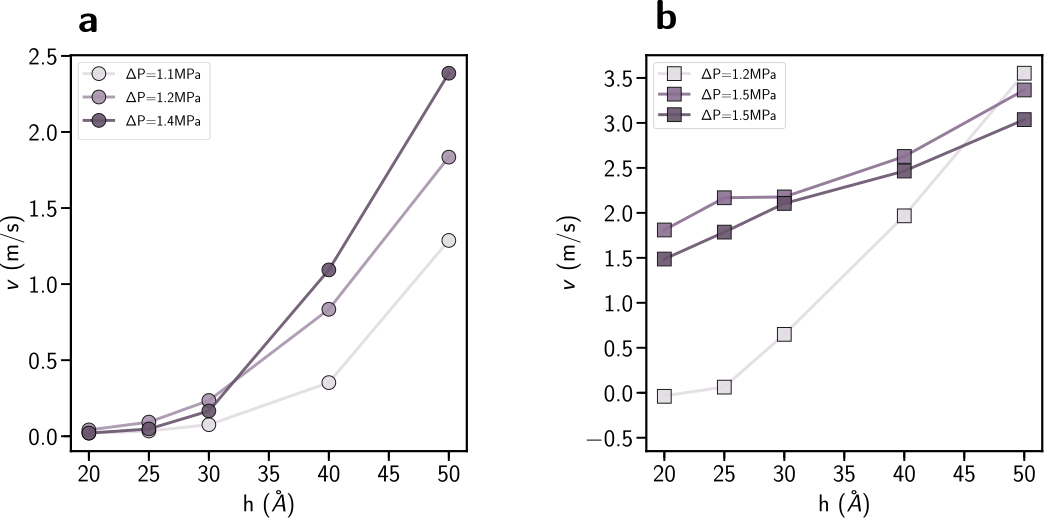}
\end{center}
\caption{\label{sifig6} a. Velocity of methane molecules in a gas state (300K and $\Delta P \sim$ 1Mpa), as a function of pore diameters in a system with constrictions. b. Same than (a) without constrictions.}
\end{figure}

\begin{figure}
\begin{center}
\includegraphics[scale=0.4]{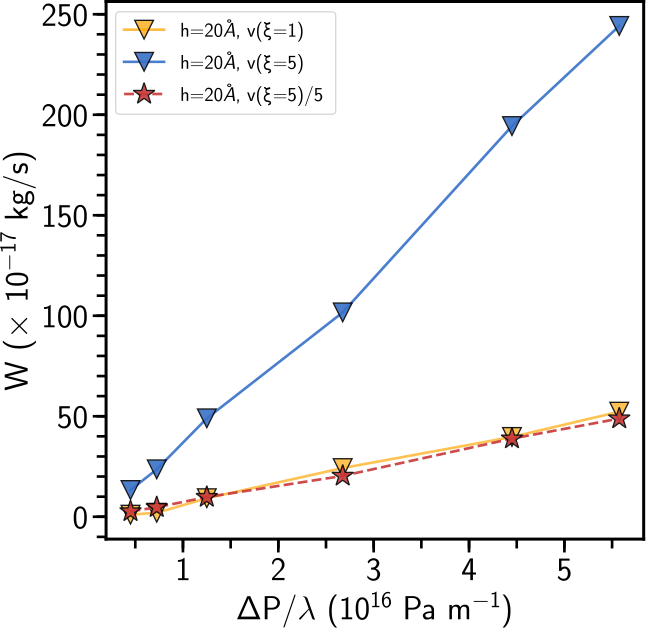}
\end{center}
\caption{\label{sifig7} Mass flowrate $W$ as a function of pressure gradient for one pore connected to the chamber $\xi=1$ (yellow triangles) and 5 pores connected to the chamber $\xi=5$ (blue triangles) with in both case a pore diameter $h=20$\AA. The last plot corresponds to the mass flowrate in a $\xi=5$ structure, normalized by the number of connections (red stars).}
\end{figure}

\clearpage
\providecommand{\noopsort}[1]{}\providecommand{\singleletter}[1]{#1}%